\documentclass{article}
\usepackage{spconf,amsmath,amssymb,graphicx,verbatim,amsfonts,xcolor,bm,multirow,subfigure}
\usepackage{algorithm,setspace}
\usepackage{hyperref}
\usepackage[noend]{algpseudocode}
\usepackage[normalem]{ulem}
\input{def.set}


\title{Boosted Locality Sensitive Hashing: \\
Discriminative Binary Codes for Source Separation}

\name{Sunwoo Kim, Haici Yang, Minje Kim\thanks{This material is based upon work supported by the National Science Foundation under Award Number:1909509.}}
\address{Indiana University\\
    Department of Intelligent Systems Engineering\\
    Bloomington, IN 47408\\
    {\small \texttt{kimsunw@indiana.edu},~ \texttt{hy17@iu.edu}, ~\texttt{minje@indiana.edu}}
}

\begin{document}
\ninept

\maketitle

\begin{abstract}
Speech enhancement tasks have seen significant improvements with the advance of deep learning technology, but with the cost of increased computational complexity. In this study, we propose an adaptive boosting approach to learning locality sensitive hash codes, which represent audio spectra efficiently. We use the learned hash codes for single-channel speech denoising tasks as an alternative to a complex machine learning model, particularly to address the resource-constrained environments. Our adaptive boosting algorithm learns simple logistic regressors as the weak learners. Once trained, their binary classification results transform each spectrum of test noisy speech into a bit string. Simple bitwise operations calculate Hamming distance to find the $\calK$-nearest matching frames in the dictionary of training noisy speech spectra, whose associated ideal binary masks are averaged to estimate the denoising mask for that test mixture. Our proposed learning algorithm differs from AdaBoost in the sense that the projections are trained to minimize the distances between the self-similarity matrix of the hash codes and that of the original spectra, rather than the misclassification rate. We evaluate our discriminative hash codes on the TIMIT corpus with various noise types, and show comparative performance to deep learning methods in terms of denoising performance and complexity. 
\end{abstract}


\section{Introduction}
Deep learning-based source separation models have improved the single-channel speech denoising performance, nearing the ideal ratio masking results (IRM) \cite{narayanan2013ideal, wang2018supervised}, or sometimes exceeding them \cite{zhen2019efficient}.  However, as the model size gets bigger, challenges grow for deploying such large models into resource-constrained devices, even just for feedforward inferences. Solving this issue is critical in real life scenarios for devices that require speaker separation and noise cancellation for quality speech communication. 

There is an ongoing research in considering the tradeoff between complexity and accuracy, which is a priority for mobile and embedded applications. Optimization of convolutional operations have been explored 
to build small, low latency models \cite{howard2017mobilenets,wang2017factorized}. Pruning less active weights \cite{HanS2016iclr} and filters \cite{li2016pruning} is another popular approach to reducing the network complexity, too. The other dimension of network compression is to reduce the number of bits to represent the network parameters, sometimes down to just one bit \cite{RastegariMCoRR16,SoudryD2014nips}
, one of its kind has shown promising performances in speech denoising \cite{KimMJ2018icassp}. 


In this paper we take another route to source separation by redefining the problem as a $\calK$-nearest neighborhood ($\calK$NN) search task: for a given test mixture, the separation is done by finding the nearest mixture spectra in the training set, and consequently their corresponding ideal binary mask vectors. However, the complexity of the search process linearly increases with the size of the training data. We expedite this tedious process by converting the query and database spectra into a hash code to exploit bitwise matching operations. To this end, we start from locality sensitive hashing (LSH), which is to construct hash functions such that similar data points are more probable to collide in the hashed space, or, in other words, more similar in terms of Hamming distance \cite{IndykP1998stoc, datar2004locality}. While simple and effective, the random projection-based nature of the LSH process is not trainable, thus limiting its performance when one uses it for a specific problem. 

We propose a learnable, but still projection-based hash function, Boosted LSH (BLSH), so that the separation is done in the binary space learned in a data-driven way \cite{li2011learning}.
BLSH reduce the redundancy in the randomly generated LSH codes by relaxing the independence assumption among the projection vectors and learn them sequentially in a {\em boosting} manner such that they complement one another, an idea shown in search applications \cite{liu2017boosting,xu2011complementary}.
 BLSH learns a set of linear classifiers (i.e. perceptrons), whose binary classification results serve as a hash code. To learn the sequence of binary classifiers we employ the adaptive boosting (AdaBoost) technique \cite{FreundY1996adaboost}, while we redefine the original classification-based AdaBoost algorithm so that it works on our hashing problem for separation. Since the binary representation is to improve the quality of the hash code-based $\calK$NN search during the separation, the objective of our training algorithm is to maximize the representativeness of the hash codes by minimizing the loss between the two self-similarity matrices (SSM) constructed from the original spectra and from the hash codes.

We evaluate BLSH on the single-channel denoising task and empirically show that with respect to the efficiency, our system compares favorably to deep learning architectures and generalizes well over unseen speakers and noises. Since binary codes can be cheaply stored and the $\calK$NN search is expedited with bitwise operations, we believe this to be a good alternative for the speaker enhancement task where efficiency matters. 



BLSH can be seen as an embedding technique with a strong constraint that the embedding has to be binary. Finding embeddings that preserve the semantic similarity is a popular goal in many disciplines. In natural language processing, Word2Vec \cite{MikolovT2013efficient}
or GloVe \cite{PenningtonJ2014glove} methods use pairwise metric learning to retrieve a distributed contextual representation that retains complex syntactic and semantic relationships within documents. Another model that trains on similarity information is the Siamese networks \cite{KochG2015oneshot,BromleyJ1994nips}
which learn to discriminate a pair of examples. Utilizing similarity information has also been explored in the source separation community by posing denoising as a segmentation problem in the time-frequency plane with an assumption that the affinities between time-frequency regions could condense complex auditory features together \cite{BachF2006learning}. Inspired by studies of perceptual grouping \cite{CookeM2001auditory}, in  \cite{BachF2006learning} local affinity matrices were constructed out of cues specific to that of speech. Then, spectral clustering segments the weighted combination of similarity matrices to unmix speech mixtures. On the other hand, deep clustering learned a neural network encoder that produces discriminant spectrogram embeddings, whose objective is to approximate the ideal pairwise affinity matrix induced from ideal binary masks (IBM) \cite{HersheyJ2016icassp}. ChimeraNet extended the work by utilizing deep clustering as a regularizer for TF-mask approximation \cite{LuoY2017chimeranet}.

\begin{algorithm}[t]
\caption{$\mathcal{K}$NN source separation}\label{alg:knn}
\begin{algorithmic}[1]
\State Input: $\bx$, $\bH$, $\bY$ \Comment{A test mixture vector and the dictionary}
\State Output: $\hat{\by}$ \Comment{A denoising mask vector}
\State Initialize an empty set $\calN=$
$\varnothing$ 
and $\mathcal{A}_\text{min}=0$
\For{$t \leftarrow 1$ to $T$}
    \If{$\calS_{\text{cos}}(\bx, \bH_{t,:})>\mathcal{A}_\text{min}$}
        \State{Replace the farthest neighbor index in $\calN$ with $t$}
        \State{Update $\mathcal{A}_\text{min}\leftarrow \min_{k\in \calN} \calS_{\text{cos}}(\bx, \bH_{k,:})$}
    \EndIf 
\EndFor
\State{return $\hat{\by} \leftarrow \frac{1}{\calK} \sum_{k \in \mathcal{N}}\bY_{k,:}$}
\end{algorithmic}
\end{algorithm}

\section{The $\calK$NN Search-Based Source Separation}


\subsection{Baseline 1: Direct Spectral Matching}
Suppose a masking-based source separation method by maintaining a large dictionary of training examples and searching for only $\calK$NN to infer the mask. We assume that if the mixture frames are similar, so are the sources in the mixture as well as their IBMs. 


Let $\bH \in \mathbb{R}^{T\times D}$ be the normalized feature vectors from $T$ frames of training mixture examples, e.g., noisy speech spectra. $T$ can be a potentially very large number as it exponentially grows with the number of sources. Out of many potential choices, we are interested in short-time Fourier transform (STFT) and mel-spectra as the feature vectors. For example, if $\bH$ is from STFT on the training mixture signals, $D$ corresponds to the number of subbands $F$ in each spectrum, while for mel-spectra $D<F$. $\bH$ is normalized with the L2 norm. We also prepare their corresponding IBM matrix, $\bY \in \{0,1\}^{T\times F}$, whose dimension $F$ matches that of STFT. For a test mixture spectrum out of STFT, $\bar{\bx}\in\Complex^{F}$, our goal is to estimate a denoising mask, $\hat{\by} \in \mathbb{R}^{F}$, to recover the source by masking, $\hat{\by}\odot\bar{\bx}$. While masking is applied to the complex STFT spectrum $\bar{\bx}$, the $\calK$NN search can be done in the $D$-dimensional feature space $\bx\in\Real^D$, e.g., $D<F$ for mel-spectra. 

Algorithm \ref{alg:knn} describes the $\calK$NN source separation procedure. We use notation $\mathcal{S}_{\text{cos}}$ as the affinity function, e.g., the cosine similarity function. For each frame $\bx$ in the mixture signal, we find the $\mathcal{K}$ closest frames in the dictionary (line 4 to 7), which forms the set of indices of $\calK$NN, $\calN=\{\tau_1, \tau_2, \cdots, \tau_\calK\}$. Using them, we find the corresponding IBM vectors from $\bY$ and take their average (line 8). 

\textbf{Complexity:} The search procedure requires a linear scan of all real-valued feature vectors in $\bH$, giving $O(QDT)$, where $Q$ stands for the floating-point precision (e.g., $Q=64$ for double precision). This procedure is restrictive since $T$ needs to be large for quality source separation. 

\begin{figure}
    \centering
    \vspace{-0.125in}
    \includegraphics[width=.91\columnwidth]{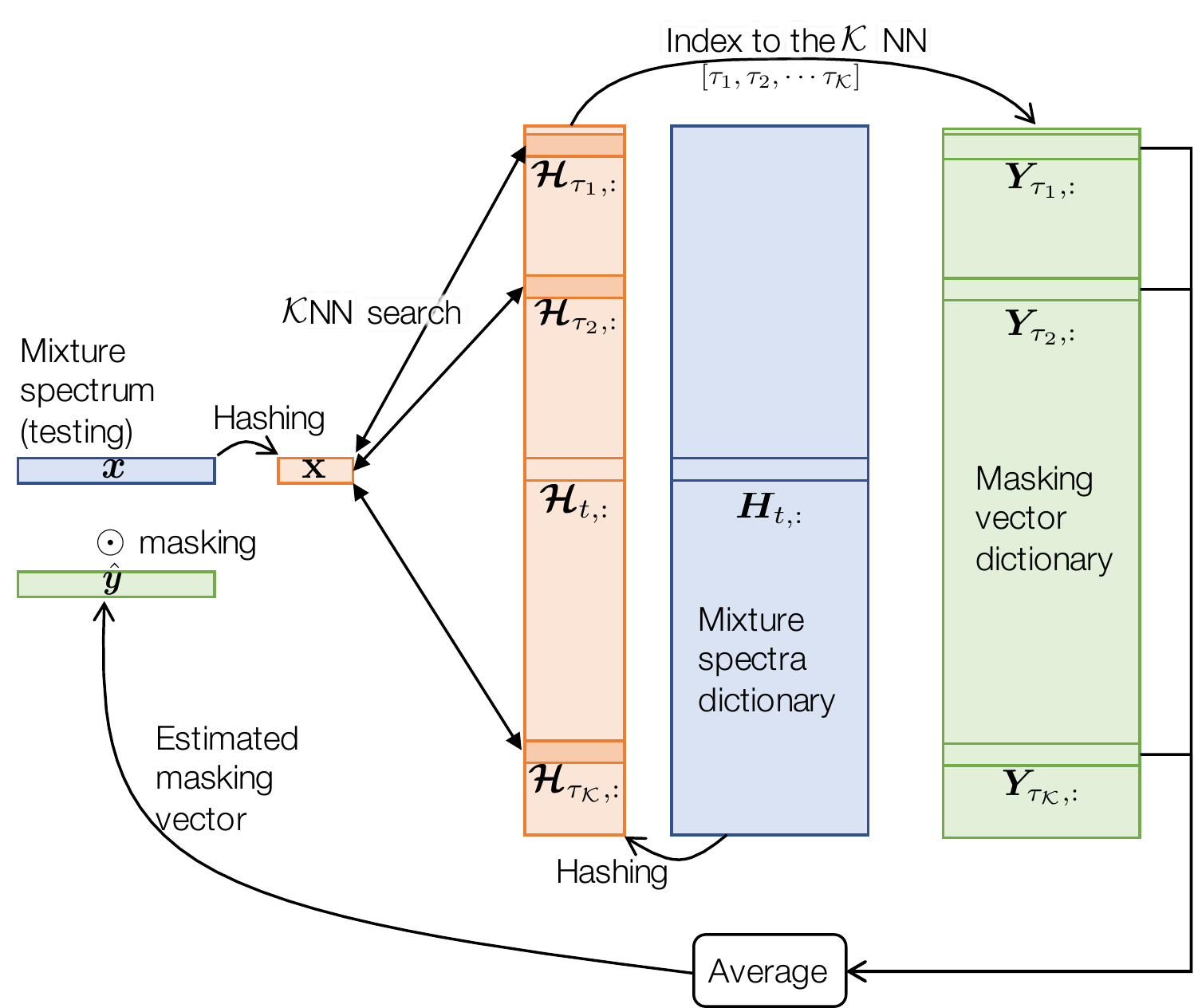}
    \caption{The $\mathcal{K}$NN-based source separation process using LSH codes.}
    \label{fig:system}
\end{figure}

\subsection{Baseline 2: LSH with Random Projections}\label{sec:lshknn}


We can reduce the storage overhead and expedite Algorithm \ref{alg:knn} using hashed spectra and the Hamming similarity between them. We define $L$ random projection vectors as $\bP \in \mathbb{R}^{L\times D}$, where the $l$-th bits in the codes is in the form of
\begin{equation}\label{eq:projection}
\bcalH_{:,l} = \text{sgn}(\bH\bP_{l,:}^\top + b_l)
\end{equation}
with $b_l$ as a bias term. Applying the same $\bP$ onto $\bH$ and $\bx$, we obtain bipolar binary $\bcalH \in \{-1, +1\}^{T\times L}$ and $\mathbf{x} \in \{-1, +1\}^{L}$, respectively. Now, we tweak Algorithm \ref{alg:knn}, by having them take the binary feature vectors. Moreover, Hamming similarity also replaces the similarity function, which counts the number of matching bits in the pair of binary feature vectors: $\calS_\text{Ham}(\ba, \bb)=\sum_l \calI (a_l, b_l)/L$, where $\calI(x,y)=1 \text{ iff } x=y$. Otherwise, the algorithm is the same. Fig. \ref{fig:system} overviews the separation process of the baseline 2. 

Randomness in the projections, however, dampens the quality of the hash bits, necessitating for longer codes. This is detrimental in terms of query time, computational cost of projecting the query to hash codes, and storage overhead from the large number of projections. The lackluster quality of the codes originates from the data-blind nature of random projections \cite{salakhutdinov2009semantic, weiss2009spectral2}.
BLSH addresses this issue by {\em learning} each projection as a weak binary classifier. 

\textbf{Complexity:} Since the same Algorithm \ref{alg:knn} is used, the dependency to $T$ remains the same, while we can reduce the complexity from $O(QDT)$ to $O(LT)$ if $L<QD$. The procedure can be significantly accelerated with supporting hardware as the Hamming similarity calculation is done through bitwise AND and pop counting. 


\begin{algorithm}[t]
\caption{BLSH training}\label{alg:BLSH}
\begin{algorithmic}[1]
\State Input: $\bH$ 
\Comment{Dictionary of training examples}
\State Output: $\bP$ \Comment{Set of projections}
\State{$\bW \leftarrow$ uniform vector of $\frac{1}{T\times T}$}
\Comment{$\bW \in \mathbb{R}^{L\times T \times T}$}
\State{$\bP \leftarrow$ random initialization}
\Comment{$\bP \in \mathbb{R}^{L\times D}$}
\State{$\beta \leftarrow$ vector of zeros}
\Comment{$\beta \in \mathbb{R}^{L}$}
\For{$l \leftarrow 1$ to $L$}
    \State{$\bP_l \leftarrow \min\limits_{\bP_l}$  $\sum\limits_{t_1, t_2} \calD\Big(\big[\bcalH_{:,l}\bcalH^{\top}_{:,l}\big]_{t_1, t_2}, \; \big[\bH\bH^\top\big]_{t_1, t_2}\Big)\bW_{l,t_1,t_2}$}
    \State{$\varepsilon_l=\sum\limits_{t_1, t_2} \calD\Big(\big[\bcalH_{:,l}\bcalH^{\top}_{:,l}\big]_{t_1, t_2}, \; \big[\bH\bH^\top\big]_{t_1, t_2}\Big)\bW_{l,t_1,t_2}$}
    \State{$\beta_l \leftarrow \frac{1}{2} \text{ln} \frac{1-\varepsilon_l}{\varepsilon_l}$}
    \State{$\bW_{l,:,:} = \bW_{l-1,:,:}\odot \exp\big(\beta_l \calD(\bcalH_{:,l-1}\bcalH^{\top}_{:,l-1}, \; \bH\bH^\top)\big)$}
\EndFor
\State{return $\bP$}
\end{algorithmic}
\end{algorithm}

\section{The proposed Boosted LSH training algorithm for Source Separation}

We set up our optimization problem with an objective to enrich the quality of the hash codes. Since the code vector is a collection of the binary classification results, during training the projection vectors are directed to minimize the discrepancies between the pairwise affinity relationships among the original spectra in $\bH$ and those we construct from their corresponding hash codes $\bcalH$.


Hence, given $\bcalH_{:,l}$ hash string from the $l$-th projection, we express the loss function in terms of the dissimilarity between the self-similarity matrices (SSM) as follows:
\begin{equation}\label{eq:xentloss}
\sum_{t_1, t_2} \calD\Big(\big[\bcalH_{:,l}\bcalH^{\top}_{:,l}\big]_{t_1, t_2}, \; \big[\bH\bH^\top\big]_{t_1, t_2}\Big)
\end{equation}
for a given distance metric $\calD$, such as element-wise cross-entropy. We scale the bipolar binary SSM to ranges of the ground truth's such that $\bcalH_{:,l}\bcalH^{\top}_{:,l} \in \{0,1\}^{T\times T}$. With this objective the learned binary hash codes can be more compact and representative than the ones from a random projection. There can be potentially many different solutions to this optimization problem, such as solving this optimization directly for the set of projection vectors $\bP$ or spectral hashing that learns the hash codes directly with no assumed projection process \cite{weiss2009spectral2}. The proposed BLSH algorithm employs an adaptive boosting mechnism to learn the projection vectors one by one.


We reformulate AdaBoost \cite{FreundY1996adaboost}, an adaptive boosting strategy for classification, which is to learn efficient weak learners in the form of linear classifiers that complement those learned in previous iterations. It is an adaptive basis function model whose weak learners form a weighted sum to achieve the final prediction, in our case an approximation of the original self-similarity that minimizes the total error as follows:
\begin{equation}\label{eq:adaboost}
\sum_{t_1, t_2} \calD \bigg(\Big[\sum_{l=1}^L \beta_l \bcalH_{:,l}\bcalH_{:,l}^\top\Big]_{t_1,t_2}, \;\;\Big[\bH\bH^\top\Big]_{t_1,t_2}\bigg),
\end{equation}
where $\beta_l$ is the weight for the $l$-th weak learner. 

\begin{figure}
\vspace{-0.1in}
\centering     
\subfigure[$L=5$]{\label{fig:a}\includegraphics[width=0.41\columnwidth]{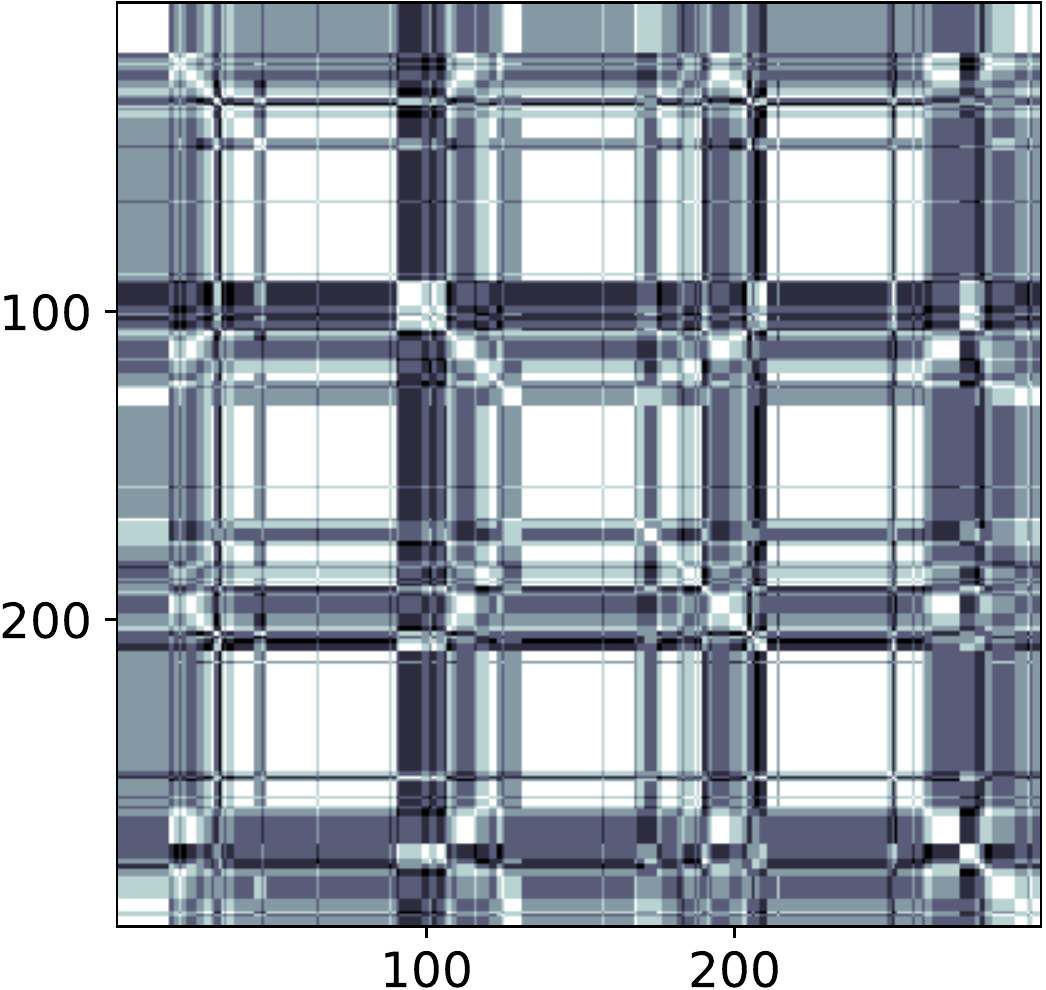}}
\subfigure[$L=50$]{\label{fig:b}\includegraphics[width=0.41\columnwidth]{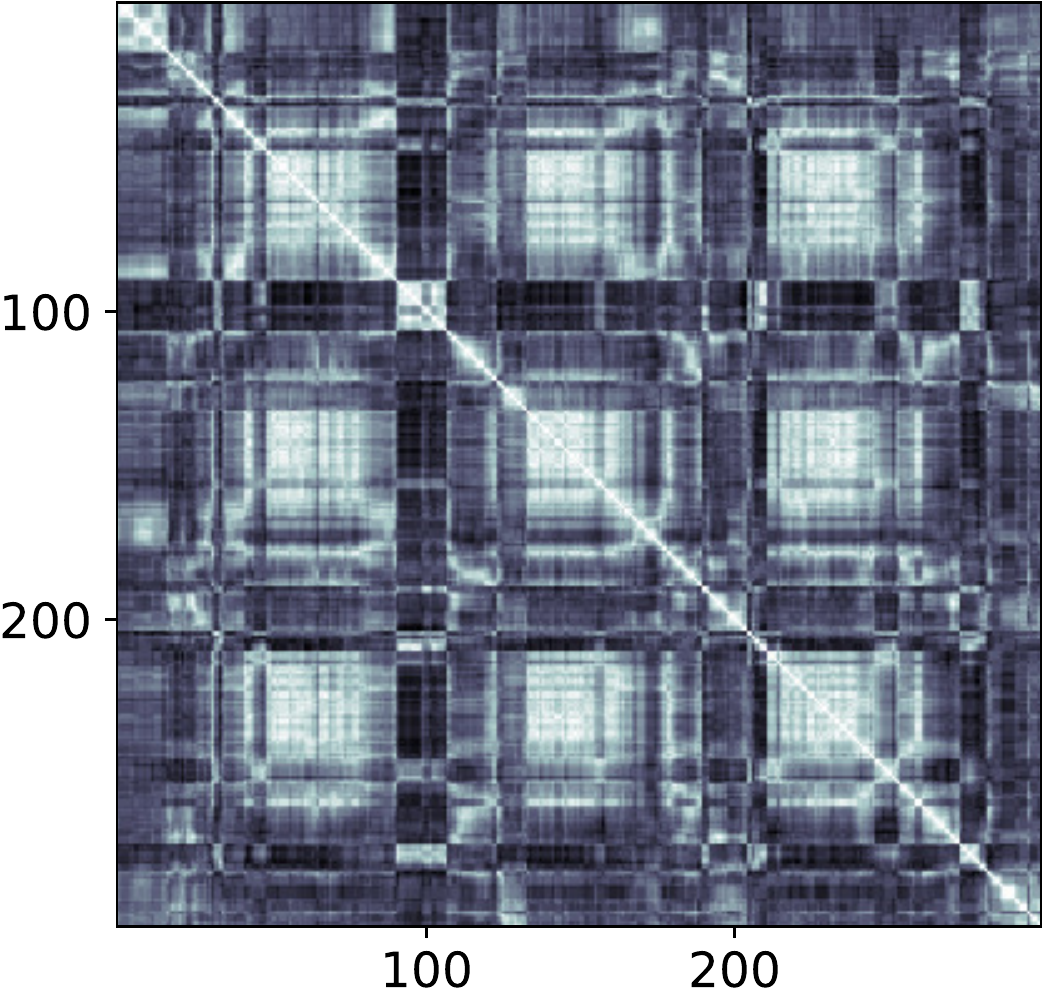}}
\subfigure[$L=200$]{\label{fig:c}\includegraphics[width=0.41\columnwidth]{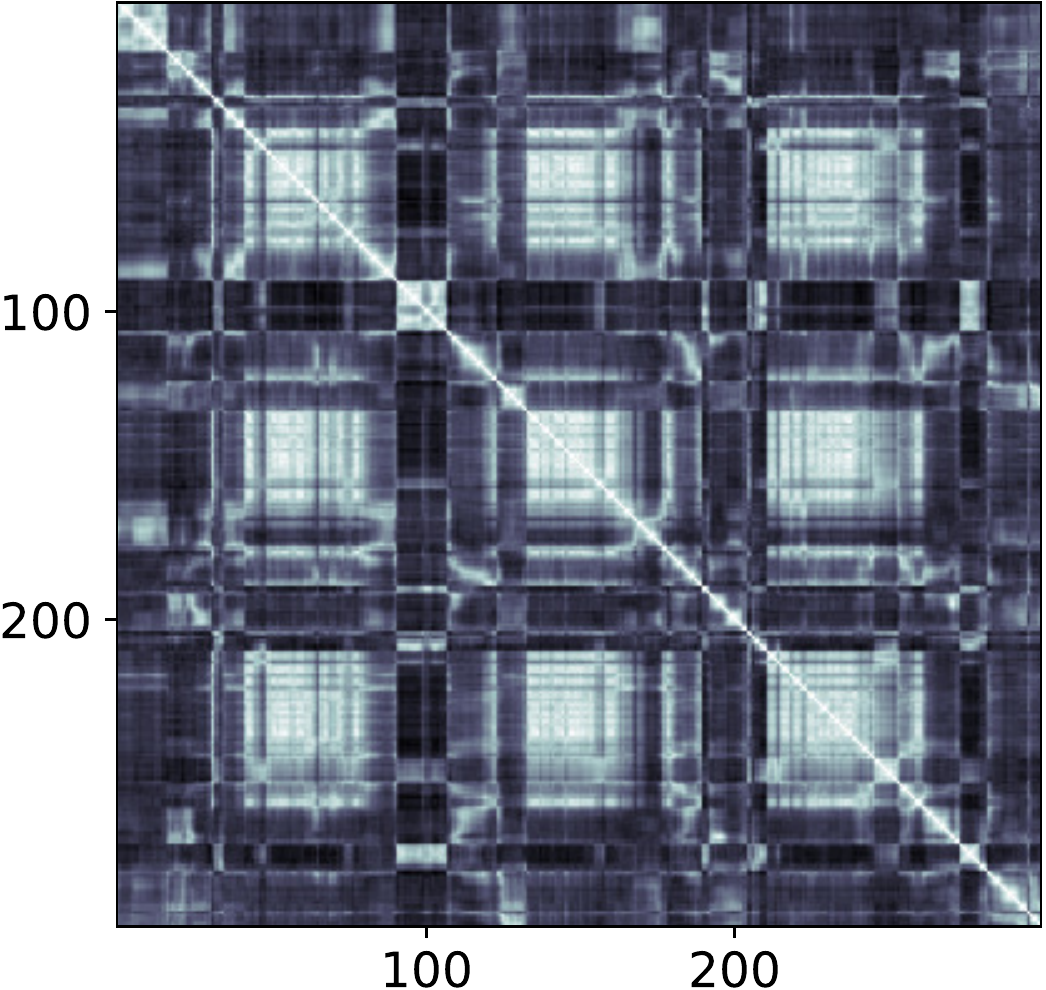}}
\subfigure[Ground Truth SSM]{\label{fig:d}\includegraphics[width=0.41\columnwidth]{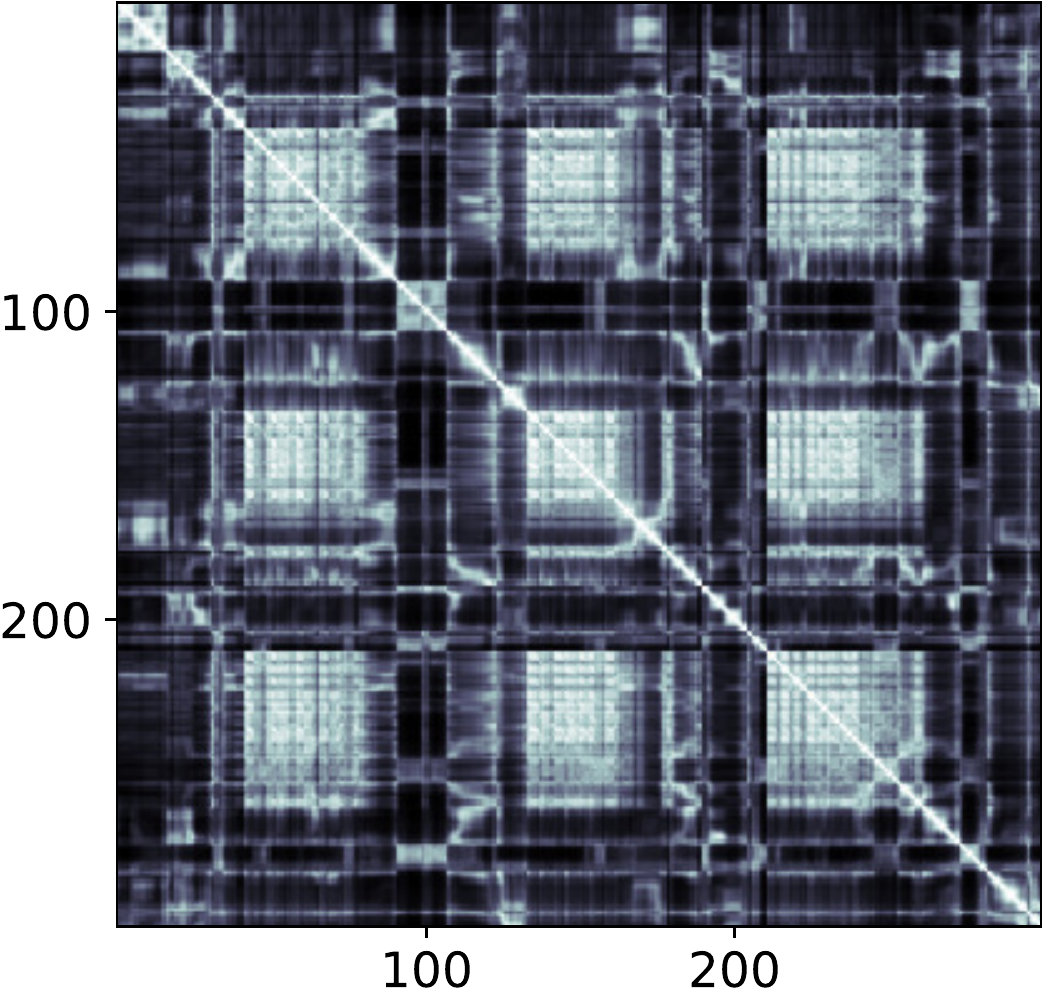}}
\caption{Self-affinity matrices of varying $L$ hash codes and original time-frequency bins}
\label{fig:bssms}
\end{figure}

In AdaBoost, the $l$-th projection is trained to focus more on the previously misclassified examples by assigning an exponentially larger weight to them. We redesign this part by making the sample weights exponentially large for the too different pairs of SSM elements as in \eqref{eq:xentloss}. The SSM weights $\bW\in\mathbb{R}^{L\times T \times T}$ are initialized with uniform values over all $(t_1, t_2)$ pairs, and then updated after adding every projection, such that each pairwise location in the SSM is weighted element-wise based on the exponentiated discrepancy made by the current weak learner as follows:  
\begin{equation}\label{eq:wupdate}
\bW_{l,:,:} = \bW_{l-1,:,:}\odot \exp\big(\beta_l \calD(\bcalH_{:,l-1}\bcalH^{\top}_{:,l-1}, \; \bH\bH^\top)\big)
\end{equation}
The effect of this update rule is to tune the weights with respect to the distances between the bitwise and original SSMs on a given metric. Weights of the elements with the largest distance computed from the $(l-1)$-th projection are amplified, and vice versa on close pairs. The $l$-th weak learner can then use these weights to concentrate on \textit{harder} examples. This approach pursues a complementary $l$-th projection; thus, overall rapidly increasing the approximation quality in \eqref{eq:adaboost} with relatively smaller $L$ projections. The final boosted objective for the $l$-th projection is formulated as
\begin{equation}\label{eq:newxentloss}
\varepsilon_l=\sum_{t_1, t_2} \calD\Big(\big[\bcalH_{:,l}\bcalH^{\top}_{:,l}\big]_{t_1, t_2}, \; \big[\bH\bH^\top\big]_{t_1, t_2}\Big)\bW_{l,t_1,t_2}
\end{equation}
Under this objective, the first \textit{few} projections index the majority of the original features, thereby dramatically reducing the overall storage of projections, computation from projecting elements, and the length of hashed bit strings. Given the learned $l$-th projection and sample weights, we obtain the weights over the projections as
\begin{equation}
\beta_l = \frac{1}{2} \text{ln} \frac{1-\varepsilon_l}{\varepsilon_l}
\end{equation}

Algorithm \ref{alg:BLSH} summarizes the procedure. Fig. \ref{fig:bssms} shows the complementary nature of the projections and their convergence behavior. After learning the projection matrix $\bP$, the rest of the test-time source separation process is the same with Algorithm \ref{alg:knn}. 

\textbf{Complexity:} BLSH does not change the run-time complexity $O(LT)$ of baseline 2. However, we expect that BLSH can outperform LSH with smaller $L$ thanks to the boosting mechanism.


\begin{figure}[!tbp]
    \centering
{\includegraphics[width=.9\columnwidth]{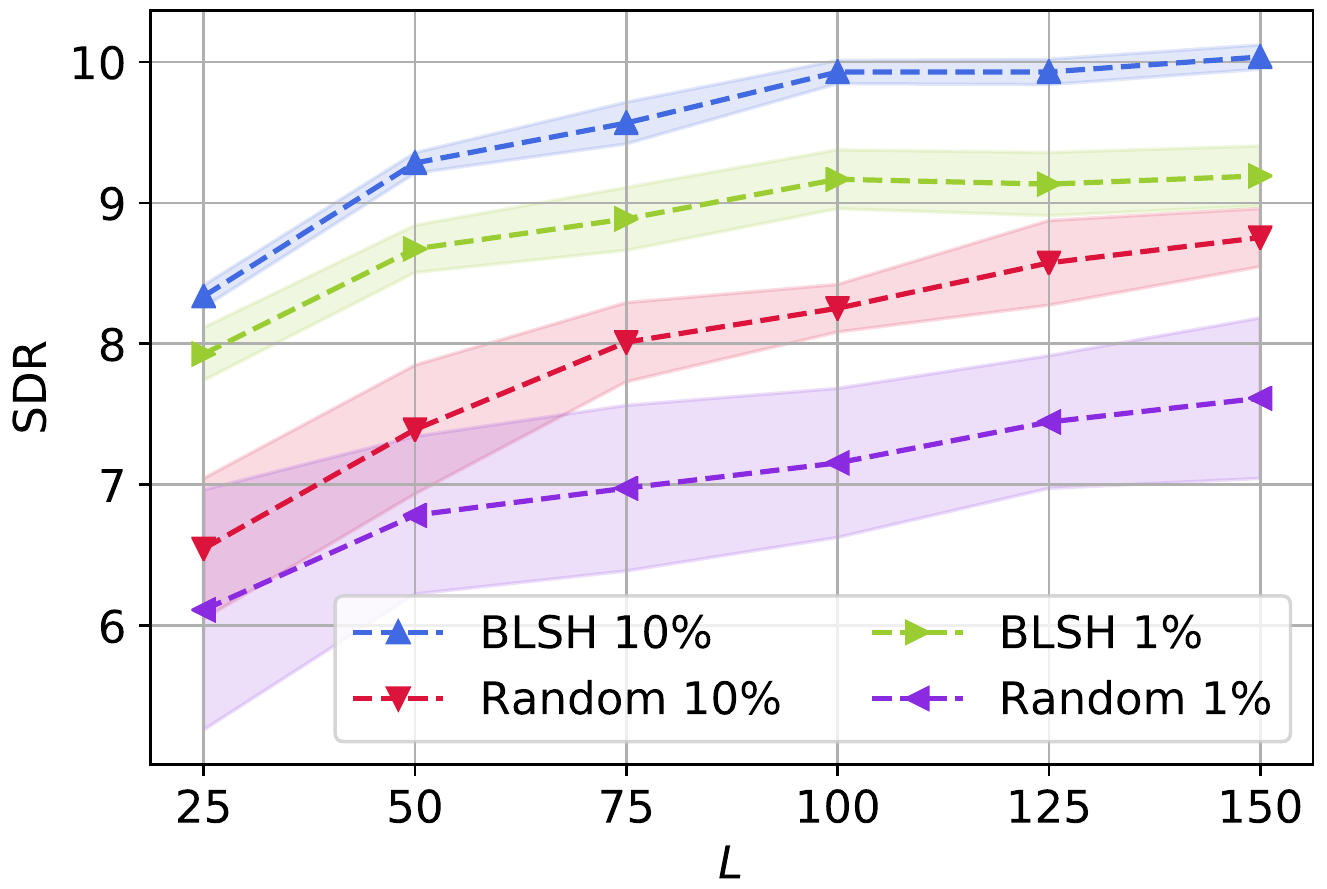}}
  \caption{SDR over number of projections on the open set.}
  \label{fig:sdrvsnproj}
\end{figure}

\section{Experiments}

\subsection{Experimental Setup}
To investigate the effectiveness of BLSH for the speech enhancement job, we evaluate our model on the TIMIT dataset. A training set consisting of 10 hours of noisy utterances was generated by randomly selecting 160 speakers from the train set, and by mixing each utterance with different non-stationary noise signals with 0 dB signal-to-noise ratio (SNR), namely $\{$birds, casino, cicadas, computer keyboard, eating chips, frogs, jungle, machine guns, motorcycles, ocean$\}$ \cite{duan2012online}. For both subsamples, we select half of the speakers as male and the other half as female. There are 10 short utterances per speaker recorded with a 16kHz sampling rate. The projections are learned on the training set and the output hash codes saved as the dictionary. 2 hours of cross validation set were generated similarly from the training set, which is used to evaluate the source separation performance of the closed speaker and noise experiments (\textbf{closed mixture set}). One hour of evaluation data set was mixed from 110 unseen speakers in the TIMIT test set and unseen noise sources from the DEMAND database, which consists of domestic, office, indoor public places, transportation, nature, and street (\textbf{open mixture set}) \cite{thiemann2013diverse}. The test set mixtures are with a 0 dB SNR and gender-balanced. We apply a Short-Time Fourier Transform (STFT) with a Hann window of 1024 and hop size of 256. During projection training, the mixture speech was segmented with the length of 1000 frames as a minibatch, such that the self-similarity matrices are of reasonable size. For evaluation, calculate the SDR, SIR, and SAR from BSS\texttt{\char`_}Eval toolbox \cite{vincent2006performance}.

\subsection{Discussion}

Fig. \ref{fig:sdrvsnproj} shows the improvements in performance over the number of projections on the open mixture set. For our evaluation, all $\calK$NN-based approaches were performed with $\calK=10$ to narrow the focus on BLSH behavior with respect to number of projections. We compare two sub-sampled dictionaries, 10\% and 1\%, respectively, to expedite the $\calK$NN search process and to reduce the spatial requirement of the hashed dictionary. We repeat the test ten times on different randomly sampled subsets. Fig. \ref{fig:sdrvsnproj} shows the average and respective standard deviations (shades). First of all,  we see that larger dictionaries with 10\% of data generally outperforms smaller dictionaries. Enlarging the length of the hash code $L$ also generally improves the performance until it saturates. More importantly, BLSH consistently outperforms the random projection method even with less data utilization (BLSH learned from 1\% of data always outperforms LSH with 10\% subset), showcasing the merit of the boosting mechanism. Also, random projections show a noticeable sensitivity to the randomness in the sub-sampling procedure: the standard deviation is significantly larger especially in the 1\% case, whereas that of BLSH remains tighter.

For comparison, we show the results in Table \ref{tab:comp} at  $L=150$ in the 10\% case. In the table, the proposed BLSH method consistently outperforms the random projection approach and also $\calK$NN on the raw STFT magnitudes in the open mixture case. 
Although the oracle IRM was higher in the open case, all systems perform worse than in the closed case.  It suggests that the unseen DEMAND noise sources are actually easier to separate, yet difficult to generalize for models trained on mixtures with ten noise types. In particular, we found that for unseen noise types, such as public cafeteria sounds, the $\calK$NN systems on the raw STFT magnitudes generalize poorly. Regardless, the proposed BLSH shows more robust generalization for the open mixture set than both the random projection method and the direct $\calK$NN on the raw STFT case. It demonstrates the better representativeness of the learned hash codes than the raw Fourier coefficients or the basic LSH codes. 

We compare the BLSH system with some deep neural network models as well, a 3-layer fully connected network (FC) and a bi-directional long short-term memory (LSTM) network \cite{schuster1997bidirectional} with two layers. Both networks are with 1024 hidden units per layer. Unsurprisingly, both networks outperform BLSH. However, considering the significantly the faster bitwise inference process of BLSH, 
more than 10dB SDR improvement is still a promising performance. 

\begin{table}[t!]
\vspace{-0.1in}
\centering
\hfill
{\caption{Evaluation metrics for different separation methods.}
\label{tab:comp}{
\setlength{\tabcolsep}{0.45em}
\begin{tabular}{c | r r | r r |r r} 
      \hline
  \multirow{2}{*}{Method}& \multicolumn{2}{c|}{SDR} & \multicolumn{2}{c|}{SAR} & \multicolumn{2}{c}{SIR}\\
  
   & Closed & Open & Closed & Open & Closed & Open \\ 
\hline
Oracle & 15.79 & 21.92 & 17.19 & 23.13 & 25.22 & 28.48 \\
\hline
BLSH  & 10.20 & 10.04 & 11.57 & 12.12 & 17.22 & 18.20\\

Random & 9.69 & 8.75 & 11.14 & 11.31 & 16.60 & 15.02\\

$\calK$NN (STFT) & 10.72 & 9.66 & 11.76 & 12.11 & 18.77 & 16.81 \\ 
 
\hline
 FC & 12.17 & 12.04 & 13.39 & 13.67 & 19.37 & 21.22\\
 LSTM & 13.37 & 12.65 & 14.79 & 14.31 & 19.68 & 21.77 \\ 
 
\hline
\end{tabular}
}}
\end{table}

Given these preliminary results, we believe further improvements can come from several areas. For example, by combining each $\beta_l$ values with the computed Hamming similarity from the $l$-th hash codes, the separation could take into account the relative contributions of the learned projections. In addition, our approach does not employ information on the time axis since each projections are applied on a frame-by-frame basis. This could be enhanced with using projections on multiple frames and windowing. Finally, enriching the dictionary to include more disparate noise types as well as optimizing training using kernel methods are potential candidates for future work.



\section{Conclusion}
In this paper, we proposed an adaptive boosted hashing algorithm called Boosted LSH for the source separation problem using nearest neighbor search. The model trains linear classifiers sequentially under a boosting paradigm, culminating to a better approximation of the original self-similarity matrix with shorter hash strings. With learned projections, the $\calK$-nearest matching frames in the hashed dictionary to the test frame codes are found with efficient bitwise operations, and a denoising mask estimated by the average of associated IBMs. We showed through the experiments that our proposed framework achieves comparable performance against deep learning models in terms of denoising quality and complexity\footnote{The code is open-sourced on \href{https://github.com/sunwookimiub/BLSH}{https://github.com/sunwookimiub/BLSH}.}\footnote{Audio examples can be found: \href{https://saige.sice.indiana.edu/research-projects/BWSS-BLSH}{https://saige.sice.indiana.edu/research-projects/BWSS-BLSH}}.

\vfill\pagebreak

\bibliographystyle{IEEEbib}
\bibliography{main,mjkim}

\begin{thebibliography}{10}

\bibitem{narayanan2013ideal}
A.~Narayanan and D.~L. Wang,
\newblock ``Ideal ratio mask estimation using deep neural networks for robust
  speech recognition,''
\newblock in {\em 2013 IEEE International Conference on Acoustics, Speech and
  Signal Processing}. IEEE, 2013, pp. 7092--7096.

\bibitem{wang2018supervised}
D.~L. Wang and J.~Chen,
\newblock ``Supervised speech separation based on deep learning: An overview,''
\newblock {\em IEEE/ACM Transactions on Audio, Speech, and Language
  Processing}, vol. 26, no. 10, pp. 1702--1726, 2018.

\bibitem{zhen2019efficient}
K.~Zhen, M.~S. Lee, and M.~Kim,
\newblock ``Efficient context aggregation for end-to-end speech enhancement
  using a densely connected convolutional and recurrent network,''
\newblock {\em arXiv preprint arXiv:1908.06468}, 2019.

\bibitem{howard2017mobilenets}
A.~G. Howard, M.~Zhu, B.~Chen, D.~Kalenichenko, W.~Wang, T.~Weyand,
  M.~Andreetto, and H.~Adam,
\newblock ``Mobilenets: Efficient convolutional neural networks for mobile
  vision applications,''
\newblock {\em arXiv preprint arXiv:1704.04861}, 2017.

\bibitem{wang2017factorized}
M.~Wang, B.~Liu, and H.~Foroosh,
\newblock ``Factorized convolutional neural networks,''
\newblock in {\em Proceedings of the IEEE International Conference on Computer
  Vision}, 2017, pp. 545--553.

\bibitem{HanS2016iclr}
S.~Han, H.~Mao, and W.~J. Dally,
\newblock ``Deep compression: Compressing deep neural networks with pruning,
  trained quantization and {H}uffman coding,''
\newblock in {\em Proceedings of the International Conference on Learning
  Representations (ICLR)}, 2016.

\bibitem{li2016pruning}
H.~Li, A.~Kadav, I.~Durdanovic, H.~Samet, and H.~P. Graf,
\newblock ``Pruning filters for efficient convnets,''
\newblock {\em arXiv preprint arXiv:1608.08710}, 2016.

\bibitem{RastegariMCoRR16}
M.~Rastegari, V.~Ordonez, J.~Redmon, and A.~Farhadi,
\newblock ``Xnor-net: Imagenet classification using binary convolutional neural
  networks,''
\newblock {\em arXiv preprint arXiv:1603.05279}, 2016.

\bibitem{SoudryD2014nips}
D.~Soudry, I.~Hubara, and R.~Meir,
\newblock ``Expectation backpropagation: Parameter-free training of multilayer
  neural networks with continuous or discrete weights,''
\newblock in {\em Advances in Neural Information Processing Systems (NIPS)},
  2014.

\bibitem{KimMJ2018icassp}
M.~Kim and P.~Smaragdis,
\newblock ``Bitwise neural networks for efficient single-channel source
  separation,''
\newblock in {\em Proceedings of the IEEE International Conference on
  Acoustics, Speech, and Signal Processing (ICASSP)}, 2018.

\bibitem{IndykP1998stoc}
P.~Indyk and R.~Motwani,
\newblock ``Approximate nearest neighbor -- towards removing the curse of
  dimensionality,''
\newblock in {\em Proceedings of the Annual ACM Symposium on Theory of
  Computing (STOC)}, 1998, pp. 604--613.

\bibitem{datar2004locality}
M.~Datar, N.~Immorlica, P.~Indyk, and V.~S. Mirrokni,
\newblock ``Locality-sensitive hashing scheme based on p-stable
  distributions,''
\newblock in {\em Proceedings of the twentieth annual symposium on
  Computational geometry}. ACM, 2004, pp. 253--262.

\bibitem{li2011learning}
Z.~Li, H.~Ning, L.~Cao, T.~Zhang, Y.~Gong, and T.~S. Huang,
\newblock ``Learning to search efficiently in high dimensions,''
\newblock in {\em Advances in Neural Information Processing Systems}, 2011, pp.
  1710--1718.

\bibitem{liu2017boosting}
X.~Liu, C.~Deng, Y.~Mu, and Z.~Li,
\newblock ``Boosting complementary hash tables for fast nearest neighbor
  search,''
\newblock in {\em Thirty-First AAAI Conference on Artificial Intelligence},
  2017.

\bibitem{xu2011complementary}
H.~Xu, J.~Wang, Z.~Li, G.~Zeng, S.~Li, and N.~Yu,
\newblock ``Complementary hashing for approximate nearest neighbor search,''
\newblock in {\em 2011 International Conference on Computer Vision}. IEEE,
  2011, pp. 1631--1638.

\bibitem{FreundY1996adaboost}
Y.~Freund and R.~E. Schapire,
\newblock ``Experiments with a new boosting algorithm,''
\newblock in {\em icml}. Citeseer, 1996, vol.~96, pp. 148--156.

\bibitem{MikolovT2013efficient}
Tomas Mikolov, Kai Chen, Greg Corrado, and Jeffrey Dean,
\newblock ``Efficient estimation of word representations in vector space,''
\newblock {\em arXiv preprint arXiv:1301.3781}, 2013.

\bibitem{PenningtonJ2014glove}
Jeffrey Pennington, Richard Socher, and Christopher~D. Manning,
\newblock ``Glove: Global vectors for word representation,''
\newblock in {\em Empirical Methods in Natural Language Processing (EMNLP)},
  2014, pp. 1532--1543.

\bibitem{KochG2015oneshot}
Gregory Koch, Richard Zemel, and Ruslan Salakhutdinov,
\newblock ``Siamese neural networks for one-shot image recognition,''
\newblock in {\em ICML deep learning workshop}, 2015, vol.~2.

\bibitem{BromleyJ1994nips}
J.~Bromley, I.~Guyon, Y.~Le{C}un, E.~S{\"a}ckinger, and R.~Shah,
\newblock ``Signature verification using a ``siamese" time delay neural
  network,''
\newblock in {\em Advances in Neural Information Processing Systems (NIPS)},
  1994, pp. 737--744.

\bibitem{BachF2006learning}
Francis~R Bach and Michael~I Jordan,
\newblock ``Learning spectral clustering, with application to speech
  separation,''
\newblock {\em Journal of Machine Learning Research}, vol. 7, no. Oct, pp.
  1963--2001, 2006.

\bibitem{CookeM2001auditory}
Martin Cooke and Daniel~PW Ellis,
\newblock ``The auditory organization of speech and other sources in listeners
  and computational models,''
\newblock {\em Speech communication}, vol. 35, no. 3-4, pp. 141--177, 2001.

\bibitem{HersheyJ2016icassp}
John~R. Hershey, Zhuo Chen, Jonathan {Le Roux}, and Shinji Watanabe,
\newblock ``Deep clustering: Discriminative embeddings for segmentation and
  separation,''
\newblock in {\em Proceedings of the IEEE International Conference on
  Acoustics, Speech, and Signal Processing (ICASSP)}, Mar. 2016.

\bibitem{LuoY2017chimeranet}
Y.~{Luo}, Z.~{Chen}, J.~R. {Hershey}, J.~{Le Roux}, and N.~{Mesgarani},
\newblock ``Deep clustering and conventional networks for music separation:
  Stronger together,''
\newblock in {\em Proceedings of the IEEE International Conference on
  Acoustics, Speech, and Signal Processing (ICASSP)}, March 2017, pp. 61--65.

\bibitem{salakhutdinov2009semantic}
R.~Salakhutdinov and G.~Hinton,
\newblock ``Semantic hashing,''
\newblock {\em International Journal of Approximate Reasoning}, vol. 50, no. 7,
  pp. 969--978, 2009.

\bibitem{weiss2009spectral2}
Y.~Weiss, A.~Torralba, and R.~Fergus,
\newblock ``Spectral hashing,''
\newblock in {\em Advances in neural information processing systems}, 2009, pp.
  1753--1760.

\bibitem{duan2012online}
Z.~Duan, G.~J. Mysore, and P.~Smaragdis,
\newblock ``Online {PLCA} for real-time semi-supervised source separation,''
\newblock in {\em International Conference on Latent Variable Analysis and
  Signal Separation}. Springer, 2012, pp. 34--41.

\bibitem{thiemann2013diverse}
J.~Thiemann, N.~Ito, and E.~Vincent,
\newblock ``The diverse environments multi-channel acoustic noise database
  (demand): A database of multichannel environmental noise recordings,''
\newblock in {\em Proceedings of Meetings on Acoustics ICA2013}. ASA, 2013, p.
  035081.

\bibitem{vincent2006performance}
E.~Vincent, R.~Gribonval, and C.~F{\'e}votte,
\newblock ``Performance measurement in blind audio source separation,''
\newblock {\em IEEE transactions on audio, speech, and language processing},
  vol. 14, no. 4, pp. 1462--1469, 2006.

\bibitem{schuster1997bidirectional}
M.~Schuster and K.~K. Paliwal,
\newblock ``Bidirectional recurrent neural networks,''
\newblock {\em IEEE Transactions on Signal Processing}, vol. 45, no. 11, pp.
  2673--2681, 1997.

\end{thebibliography}

\end{document}